\documentstyle[epsf,preprint,prl,aps]{revtex}
\widetext
\tighten
\def\be{\begin{equation}}
\def\ee{\end{equation}}
\def\bea{\begin{eqnarray}}
\def\eea{\end{eqnarray}}

\def\bl{\biggl}
\def\br{\biggr}

\begin{document}
\draft
\title
{ de Broglie-Bohm interpretation for wave function of 
Reissner-Nordstr\"{o}m-de Sitter black hole
} 

\author{Masakatsu Kenmoku 
\footnote{Electronic address: kenmoku@cc.nara-wu.ac.jp}} 
\address{ Department of Physics, Nara Women's University, 
Nara 630, Japan}   

\author{Hiroto Kubotani 
\footnote{Electronic address:kubotani@cpsun3.b6.kanagawa-u.ac.jp}} 
\address{ Faculty of Engineering, Kanagawa University, 
Yokohama 221, Japan}
 
\author{Eiichi Takasugi 
\footnote{Electronic address:takasugi@phys.wani.osaka-u.ac.jp}}
\address{Department of Physics, Osaka University, Osaka 560, Japan }

\author{Yuki Yamazaki 
\footnote{Electronic address:yamazaki@phys.nara-wu.ac.jp}}
\address{Graduate School of Human Culture, Nara Women's University,
Nara 630, Japan}

\maketitle

\begin{abstract}

We study the canonical quantum theory of the 
Reissner-Nordstr\"{o}m-de Sitter black hole(RNdS).  
We obtain an exact general solution of the Wheeler-DeWitt equation 
for the spherically symmetric geometry with electro-magnetic field.
We investigate the wave function 
form a viewpoint of the de Broglie-Bohm interpretation. 
The de Broglie-Bohm interpretation introduces a rigid trajectory 
on the minisuperspace without assuming an outside observer 
or causing collapse of the wave function.
In our analysis, we obtain the boundary condition for the wave 
function which corresponds to the classical 
RNdS black hole and describe the quantum fluctuations near the horizons
quantitatively.   
   
\end{abstract}


\section{Introduction}
The canonical form of general relativity is presented by Dirac
\cite{Dirac64}, and by Arnowitt, Deser and Misner (ADM) \cite{ADM62}. 
In the formulation, the dynamics of the gravitational field is given as 
the totally constraint system. 
For quantization of the gravity, the constraints are used as
operator restrictions on the state, and the Hamiltonian
constraint gives the Wheeler-DeWitt(WD) equation\cite{Wheeler,DeWitt67}.  
The canonical quantum gravity based on the Wheeler-DeWitt approach 
has mathematical and conceptual difficulties. 
The WD equation is 
a functional differential equation with respect to the three 
dimensional metric components $g_{\mu\nu}$ and includes 
the product of the functional derivatives. 
When the space is inhomogeneous, one can hardly solve this equation 
of infinite degrees of freedom. 
It has also an operator ordering ambiguity.    
If one reduce the degrees of freedom to finite by symmetry, 
the WD equation becomes a greatly simple 
and solvable equation.    
Kucha$\check{\rm r}$ studied the geometrodynamics 
of Schwarzschild black holes \cite{Kuchar94}. 
The black hole mass is shown to be a dynamical variable 
on the phase space. 
Nakamura, Konno, Oshiro and Tomimatsu applied the canonical formulation 
to the inside of the horizon of the Schwarzschild black hole, 
and obtained an exact solution of the WD equation 
under a condition of a mass eigenstate equation.
In the WKB region, their solution correspond to a classical 
black hole solution
and they derived a semiclassical picture from the exact solution 
\cite{Nakamura93}. 
The WKB approximation breaks down if a variation of a potential 
is considerably large over a wave length. 
In this situation the frequency of the wave function is very high 
and the amplitude violently fluctuates. 
Therefore they were inhibited from interpreting the wave 
function near the horizon. 
In the region where the scalar curvature becomes very large 
such as a horizon, a sensible quantum effect of the spacetime is expected 
to appear prominently. 
We cannot neglect the influence of the spacetime fluctuation 
on the Hawking radiation \cite{Hawking75,Gibbons-Hawking}. 
In our preceding paper\cite{KKTY}, 
we solved the WD equation for a minisuperspace 
and obtained the wave function for the Schwarzschild black hole. 
Then we applied the de Broglie-Bohm(dBB) 
interpretation to the wave function in order to estimate the quantum 
fluctuations instead of the WKB approach. 
Further we investigated the operator ordering ambiguity 
on the quantum effect. 
In the dBB approach\cite{deBroglie,Bohm52,Bell87}, 
by introducing a rigid trajectory picture 
thorough the phase factor of the wave function, 
we can make quantitative estimations 
of quantum effects near horizons and make a natural derivation of 
the (semi)classical picture from the quantum theory without 
having to assume an outside observer and having to cause 
the collapse of the wave function.  
The emergence of the classical picture from quantum 
systems is a problem under debate by several authors \cite{Halliwell87}. 
In regard to the conceptual problem of time, that is, there is no time 
evolution in the WD equation, we can also get the 
parametric time in a natural way, although the general covariance 
spontaneously breaks down at a quantum level.  

The purpose of our work is to investigate static states of quantum
geometry near the horizons. 
We solve the WD equation  for the 
Reissner-Nordstr\"{o}m-de Sitter(RNdS) black hole.
The RNdS geometry has some specific features such that  
the Cauchy horizon and the timelike singularity exist. 
In our minisuperspace, the dBB trajectory picture shows that 
quantum effects are dominant near all horizons 
while the trajectory asymptotically 
approaches to the classical one in an infinite region 
or the singularity.

%
\section{Canonical formalism}
The general form of the Einstein-Maxwell action with the cosmological 
constant is written as  
\begin{equation}
 {\cal S}  =  \frac{1}{16\pi}\int d^4x \sqrt{-^{(4)} g} \biggl[
    ( \ ^{(4)}R-2 \lambda )-F_{\mu\nu}F^{\mu\nu} \biggr].
\label{Gravaction}
\end{equation} 
In this expression $\ ^{(4)}R$ is the four-dimensional Riemann 
curvature scalar and 
$\lambda$ denotes the cosmological constant. 
We take the natural unit $G=c=1$.  
$F_{\mu\nu}$ is the electromagnetic field strength
\begin{equation}
 F_{\mu\nu} = \partial_{\mu}A_{\nu} - \partial_{\nu}A_{\mu}.
\end{equation}
The action principle gives the Einstein-Maxwell's equation in vacuum. 
A static spherically symmetric solution of it is known as the 
Reissner-Nordstr\"{o}m-de Sitter(RNdS) metric
\begin{equation}
 ds^2 = -\biggl(1-\frac{2M}{R}+\frac{Q^2}{R^2}-\frac{\lambda}{3} R^2 \biggr)
    dT^2 + \biggl(1-\frac{2M}{R} 
    +\frac{Q^2}{R^2}
    -\frac{\lambda}{3} R^2 \biggr)^{-1} dR^2 + R^2 d\Omega^2,
\label{RNdSmetric}
\end{equation}
which describes a black hole with the mass $M$ and the electric charge 
$Q$ in the background of the static de Sitter space. 
$d\Omega^2$ denotes the line element on the unit sphere.  

We reduce the gravitational degrees of freedom by spherical symmetry anzats.
Let us restrict our attention to the inside of the horizon, 
where the radial coordinate plays the role of the time coordinate.
We denote the coordinate as $t$.
We introduce two metric variables $U$ and $V$ which depend only on $t$ as  
\begin{equation}
 ds^2 = -\frac{\alpha(t) ^2}{U(t)}~dt^2 + U(t)~ dr^2 + V(t)~ d\Omega^2,
\label{UV}
\end{equation}
where $\alpha$ is the lapse function. 
Because of spherical symmetry, the shift vector $N^a$ except 
the radial component $N^t$ must be zero. We have chosen the radial 
component $N^t$ to be zero taking account of the static spacetime.
Using the metric Eq.(\ref{UV}) the action Eq.(\ref{Gravaction}) is 
decomposed into the ADM hypersurface action   
\begin{equation}
 {\cal S}_{\Sigma} = \int dt \int dr L,
\end{equation}
where the Lagrangian is 
\begin{equation}
 L =  \frac{1}{4} \biggl(-\frac{\dot{V}\dot{U}}{\alpha}
      -\frac{U\dot{V}^2}{2\alpha V}+2\alpha \biggr)
      +\frac{V}{2\alpha} F_{01}^2 - \frac{\lambda}{2}~\alpha V,
\label{LagUV}
\end{equation} 
with
\begin{equation}
 F_{01} = \dot{A} - A~'_0,
\end{equation}
which is the $(0,1)$ component of the field strength $F_{\mu\nu}$.  
Here the notations are $\cdot = \partial/\partial t~$ 
and $~'= \partial/\partial r$.
Although we have assumed that the electromagnetic field $A_1 (\equiv A)$ 
depends only on time variable $t$, 
the field $A_0$ is a redundant degree of freedom and we need not to 
impose the symmetry on $A_0$.

The Euler-Lagrange equation for the system Eq.(\ref{LagUV}) gives 
the classical solution:    
\begin{eqnarray}
 \alpha &=& 1, \\
 U &=& -\biggl(1-\frac{2m}{\sqrt{V}}+\frac{Q^2}{V}
                            -\frac{\lambda}{3}V \biggr), 
\label{u}\\
 \sqrt{V} &=& t, \\     
\label{v} 
 VF_{01}&=&Q\alpha(=const),   
\end{eqnarray}
which correspond to the extension of 
the RNdS spacetime Eq.(\ref{RNdSmetric}).
The integration constant $m$ represents the asymptotically 
observed mass of a spherically symmetric matter.  
The point $V^{1/2}=t=0$ is a real singularity. 
There are three horizons at the values $t_1,t_2$ and $t_3$ 
for which $U=0$. 

For convenience we change the variables from $U$ and $V$ to 
$z_+$ and $z_-$ as 
\begin{equation}
 z_+ \equiv U \sqrt{V}, \hspace{0.5cm}  
 z_- \equiv   \sqrt{V}.  
\label{z}
\end{equation}
By using these new variables, the Lagrangian Eq.(\ref{LagUV}) 
becomes a simpler and symmetric form
\begin{equation}
  L = \frac{1}{2} \biggl( -\frac{1}{\alpha} \dot{z}_+ 
       \dot{z}_- + \alpha \biggr) + \frac{z_-^2}{2\alpha}F_{01}^2 
      - \frac{\lambda}{2}\alpha z_-^2.
\label{Lagz}
\end{equation}
The canonical momenta conjugate to $z_+,\ z_-$ and 
$A$ are obtained from the Lagrangian:
\begin{eqnarray}
 {\it \Pi}_+ &\equiv& \frac{\partial L}{\partial \dot{z}_+} 
     = -\frac{1}{2\alpha} \dot{z}_-, 
\label{P+cl}\\
 {\it \Pi}_- &\equiv& \frac{\partial L}{\partial \dot{z}_-} 
    = -\frac{1}{2\alpha} \dot{z}_+, 
\label{P-cl}\\
 {\it \Pi}_A &\equiv& 
      \frac{\partial L}{\partial \dot{A}} 
    = \frac{z_-^2}{\alpha}(\dot{A}-A'_0). 
\label{A}
\end{eqnarray}
Since the Lagrangian Eq.(\ref{Lagz}) do not contain the terms
$\dot{\alpha}$ and $\dot{A_0}$, 
the corresponding momenta vanish trivially and they yield 
the primary constraints. 
The Legendre transformation gives the Hamiltonian and      
the action for this system is
\begin{equation}
 {\cal S}_{\Sigma} = \int dt \int dr ( {\it\Pi}_A\dot{A} 
    + {\it\Pi}_+ \dot{z}_+  
    +  {\it\Pi}_- \dot{z}_- -\alpha H - A_{0} H_{A}),
\label{ADMaction}
\end{equation}
where $~H~$ and $~H_A~$ are
\begin{eqnarray}
 H &=& -2~ {\it\Pi}_+ ~{\it\Pi}_- - \frac{1}{2} 
       + \frac{{\it\Pi}^2_A}{2z_-^2}  +\frac{\lambda}{2} z^2_-, 
\label{Hconstraint} \\
 H_A &=& -({\it\Pi}_A)~'.
\label{HAconstraint}
\end{eqnarray}
Here we integral out over the unit sphere and treat the integral 
$\int^{\infty}_0 dr$ to be finite.
$\it\Pi_\alpha$ and $\it\Pi_{A_0}$ have vanishing Poisson brackets with 
the Hamiltonian and also generate the secondary constraints 
which give the Hamiltonian constraints: $H \approx 0$ and $H_A \approx 0$.  
The secondary constraints do not generate further constraints, 
because the Poisson brackets with the Hamiltonian weakly vanish. 
Then the Hamiltonian constraints are the first class. 
We also introduce the black hole mass as a dynamical variable 
on canonical data. Following the calculation of the Schwarzschild mass 
by Kucha$\check{\rm r}$, 
the RNdS black hole mass is expressed as 
\begin{equation}
 M = 2~ {\it\Pi}_+ ~ z_+ ~{\it\Pi}_+  + \frac{z_-}{2} 
     +\frac{{\it\Pi}^2_A}{2z_-} - \frac{\lambda}{6} z_-^3.
\label{clM}
\end{equation}

Next we quantize the black hole system in 
the Schr\"{o}dinger representation. The canonical momenta 
are quantized as usual differential operators:
\begin{equation}
\widehat{{\it\Pi}}_+ =-i\hbar \frac{\partial}{\partial z_+}, \hspace{0.5cm}
\widehat{{\it\Pi}}_- =-i\hbar \frac{\partial}{\partial z_-}, \hspace{0.5cm}
\widehat{{\it\Pi}}_A =-i\hbar \frac{\partial}{\partial A}. 
\end{equation}
We take the Weyl ordering in the following calculations.
Following Dirac's canonical quantization procedure,  
we impose operator restrictions on the state vector $\Psi$ 
as constraints.  
For the Hamiltonian constraint Eq.(\ref{Hconstraint}),  
\begin{equation}
\widehat{H} \Psi 
 = \biggl(- 2~ \widehat{{\it\Pi}}_+~ \widehat{{\it\Pi}}_- -\frac{1}{2} 
    +\frac{\widehat{\it\Pi}^2_A}{2 z^2_-} +\frac{\lambda}{2} z^2_- 
    \biggr) \Psi(z_+,z_-,A) =0,
\label{WD} 
\end{equation}
which is called the Wheeler-DeWitt(WD) equation. 
The mass operator $\widehat{M}$ is weakly commutable with 
the Hamiltonian in the Weyl ordering:  
\begin{equation}
 [ \widehat{H},~ \widehat{M}]\Psi = 2i\hbar ~ \widehat{{\it\Pi}}_+
    ~ \widehat{H}\Psi = 0.
\end{equation}
In addition to the WD equation, we also impose another two 
restrictions on the state vector $\Psi$ as constraint equations, 
the mass operator $\widehat{M}$ with the mass eigenvalue $m$ 
and the Hamiltonian $H_A$ with the eigenvalue of the charge $Q$: 
\begin{equation} 
 \widehat{M}\Psi= 
   \biggl( 2~  \widehat{{\it\Pi}}_+~ z_+~ \widehat{{\it\Pi}}_+ 
     + \frac{z_-}{2}  
   +\frac{\widehat{\it\Pi}^2_A}{2 z_-} - \frac{\lambda}{6}~ z^3_- 
   \biggr) \Psi(z_+,z_-,A) 
   = m~ \Psi(z_+,z_-,A),
\label{M}
\end{equation}
\begin{equation}
 \widehat{\it\Pi}_A~ \Psi(z_+,z_-,A)= Q~ \Psi(z_+,z_-,A).
\label{Q}
\end{equation}
Eq.(25) is equivalent to the charge conservation law
on the state vector $\Psi$. 
First we solve the equation (\ref{Q})
\begin{equation}
 \biggl( -i \hbar \frac{\partial}{\partial A} + Q \biggr) \Psi = 0.
\end{equation}
The solution is 
\begin{equation}
 \Psi = \psi(z_+,z_-)e^{-iQA/\hbar},
\end{equation}
where $\psi$ is a general function of $z_+$ and $z_-$.
Next we shall solve the two equations Eqs.(\ref{WD}) and (\ref{M}) to 
determine $\psi$. 
Instead of solving the mass eigenvalue equation directly, 
we consider an eigenvalue equation derived from the linear 
combination of the Hamiltonian and the mass operators: 
\begin{equation}
 \widehat{L}\Psi \equiv [{\it\Pi}_+ ~z_+ ~\widehat{H} + {\it\Pi}_- 
   ~(\widehat{M}-m)]\Psi = 0,
\label{L}
\end{equation}
and then, after making a variable transformation 
from $z_-$ to $\tilde{z}_-$
\begin{equation}
 \tilde{z}_- =  \frac{1}{z_-} \biggl(z_-^2 -2mz_- -Q^2 
        - \frac{\lambda}{3}z_-^4 \biggr),
\end{equation}
we can obtain the first order differential equation
\begin{equation}
 \biggl(\tilde{z}_- \frac{\partial}{\partial \tilde{z}_-} 
     - z_+ \frac{\partial}{\partial z_+} \biggr) \psi(z_+,z_-) = 0.
\end{equation}
This equation is the symmetric with respect to the variables $z_+$ 
and $\tilde{z}_-$. Then we further transform the variables $z_+$ and 
$\tilde{z}_-$ to $z$ and $y$:  
\begin{equation}
 y \equiv  \frac{1}{\hbar}\sqrt{-z_+/\tilde{z}_- },  \hspace{0.5cm}
 z \equiv  \frac{1}{\hbar}\sqrt{-z_+~\tilde{z}_- },
\label{yz}
\end{equation}
Using the variables Eq.(\ref{yz}), Eqs.(\ref{L}) 
and (\ref{WD}) become the form:
\begin{eqnarray}
 y~ \frac{\partial}{\partial y}~ \psi (y,z) &=& 0,
\\
 \biggl(~\frac{1}{z} \frac{\partial}{\partial z} 
  z \frac{\partial}{\partial z} + 1 \biggr) \psi(y,z) &=& 0.
\label{Bessel}
\end{eqnarray}
The equation (\ref{Bessel}) for $\psi$ is the Bessel's differential 
equation with zeroth order. 
Then we finally obtain an exact solution of 
the RNdS quantum black hole:  
\begin{eqnarray}
 \Psi(z,A) &=& \biggl(c_1H^{(1)}_0(z)+c_2 H^{(2)}_0(z)\biggr)e^{-iQA/\hbar},
\label{solution}\\
 z &=& \sqrt{-z_+ \tilde{z}_-} = \biggl[-u \biggl(v-2m\sqrt{v}+Q^2
       -\frac{\lambda}{3}~v^2\biggr)\biggl]^{1/2},
\label{fisolution}
\end{eqnarray}
where $c_1$ and $c_2$ are integration constants.
The Hankel functions $H^{(1)}_0$ and $H^{(2)}_0$ are 
linearly independent and complex conjugate to each other.
If the charge $Q=0$ and the cosmological constant $\lambda=0$, 
the quantum RNdS solution agrees with the quantum Schwarzschild one 
\cite{KKTY}. 
\section{de Broglie-Bohm interpretation}
%
In the de Broglie-Bohm(dBB) interpretation 
the quantum mechanics is explained as follows. 
First, the wave function  
$\Psi$ is given as the solution of the Schr\"{o}dinger equation.
In our case the corresponding equation is the Wheeler-DeWitt(WD) 
equation, which has no time evolution and 
whose eigen state vector is considered as a stationary state 
with zero energy.  
Next, the wave function is decomposed into two real functions, 
the amplitude ${\sf R}$ and the phase ${\sf S}$ according to the expression 
$\Psi(z_+,z_-,A)= {\sf R}(z_+,z_-,A)
\exp[i{\sf S}(z_+,z_-,A)/\hbar]$. 
Substituting this expression into the WD equation (\ref{WD}), 
we can rewrite it in the real part and the imaginary part equations:  
\begin{eqnarray}
   2\frac{\partial {\sf S}}{\partial z_+} 
   \frac{\partial {\sf S}}{\partial \tilde{z}_-} + \frac{1}{2} + V_Q  = 0, 
\label{reWD} \\ 
   \frac{\partial}{\partial z_+}(z_+ R^2) 
   \frac{\partial {\sf S}}{\partial z_+} =0, 
\label{imWD}
\end{eqnarray}
where the function $V_Q$ is
\begin{equation}
   V_Q = -\frac{2\hbar^2}{{\sf R}}\frac{\partial^2}
          {\partial z_+ \partial \tilde{z}_-}{\sf R}.  
\label{VQ}
\end{equation}
Here we use the constraint equation for the charge Eq.(\ref{Q}) in advance.
As a result the dependence of the function $V_Q$ and the equations 
(\ref{reWD}) and (\ref{imWD}) on the variable $A$ is removed. 

In the dBB interpretation, 
we introduce the trajectories $Z_+(T),Z_-(T)$ and $A(T)$ on which 
the particles are assumed to move with the momenta: 
\begin{eqnarray}
  {\it\Pi}_+ &=& -\frac12 \dot{Z}_- = 
\br. \frac{\partial {\sf S}}{\partial z_+}\bl|_{z_+ = Z_+, z_- = Z_-,A=A} 
\label{P+} , \\
  {\it\Pi}_- &=& -\frac12 \dot{Z}_+ = 
\br. \frac{\partial {\sf S}}{\partial z_-}\bl|_{z_+ = Z_+, z_- = Z_-,A=A}
\label{P-}, \\
 {\it\Pi}_A &=& \dot{A} = 
\br. \frac{\partial {\sf S}}{\partial A} \bl|_{z_+=Z_+,z_-=Z_-,A=A}. 
\label{PA}
\end{eqnarray}
Here $\cdot = \partial/\partial T$ is a derivative with respect to a 
parameter 
introduced through the phase factor ${\sf S}$. 
These trajectories are assumed to be a statistical ensemble of 
the probability distribution given by ${\sf R}^2$.
In this interpretation, Eqs.(\ref{reWD}) and (\ref{imWD}) 
indicate the Hamilton-Jacobi equation and 
the continuity equation of the probability, respectively.
In the Eq.(\ref{reWD}), there is a term $V_Q$ which is not present 
in the classical Hamilton-Jacobi equation.  
The trajectories are modified by this term quantitatively.
If $V_Q$ is negligible compared with the classical potential, 
the quantum trajectory approaches to the classical one and indeed 
this situation is what we call {\it classical}. 
In this sense we call the function $V_Q$ {\it quantum potential}. 
We recall that the quantum theory itself 
is applied to the system of an ensemble, 
that is, one must perform many measurements to one 
particle to obtain the wave function. 
In the ordinary Copenhagen interpretation, the system is divided into 
the external observer described by the classical mechanics and the quantum 
system, and the notion of the probability enters the theory. 
In the de Broglie-Bohm interpretation, on the other hand, 
there is no division between the classical observer 
and the quantum system and no collapse of the wave function.  
The predictability of the quantum mechanics stems from the notion of 
the distribution of a statistical ensemble of well defined quantum 
trajectories, which are modified by a quantum effect produced by the 
quantum potential. 

For our minisuperspace model, 
we calculate the quantum potential and the quantum trajectories. 
We select the second term of the solution 
Eq.(\ref{solution}) 
\begin{equation}    
   \Psi(z,A)= N H^{(2)}_0(z)e^{-iQA} \equiv {\sf R}(z,A)
   e^{i{\sf S}(z,A)/\hbar},
\label{hankel} 
\end{equation}
since $H_0^{(2)}$ is a outgoing wave from the singularity 
at the origin to the outside.  
From the dBB point of view, it will be shown to correspond 
to the classical black hole (Fig.2).
On the contrary, the superposition of $H^{(1)}$ and $H^{(2)}$ 
in Eq.(\ref{fisolution}) does not approach any classical solution.
The equations on the velocities Eqs.(\ref{P+}),(\ref{P-}) 
and (\ref{PA}) are obtained as
\begin{eqnarray}
  \dot{Z}_+ &=& \frac{2\hbar}{\pi}
         \frac{Z^2_- -\lambda Z_-^4 - Q^2}{Z^2_- Z_+ |H^{(2)}_0(Z)|^2}, 
\label{dotZ+} \\
  \dot{Z}_- &=& \frac{2\hbar}{\pi}\frac{1}{Z_+ |H^{(2)}_0(Z)|^2}, 
\label{dotZ-} \\
  \dot{A} &=& Q.
\end{eqnarray}
The ratio of Eq.(\ref{dotZ+}) to Eq.(\ref{dotZ-}) 
gives a functional relation $(Z_+,Z_-)$.
\begin{equation}
 Z_+ = c_0 \biggl(Z_- -2m+\frac{Q^2}{Z_-}-\frac{\lambda}{3}Z^3_-\biggr),
\label{+-rel}
\end{equation}
where $c_0$ is an integration constant.
With the choice of  
$c_0=-1$ this relation is translated back to that of the original 
metric variables $U$ and $V$ in Eq.(\ref{UV}):
\begin{equation}
 U = -\biggl(1-\frac{2m}{V^{1/2}} + \frac{Q^2}{V} 
    - \frac{\lambda}{3}~ V \biggr),  
\label{UVrel}
\end{equation}
which corresponds to the classical relation in Eq.(\ref{u}).  
There are three horizons corresponding to the null surfaces $U=0$.   

The quantum potential Eq.(\ref{VQ}) is expressed as  
\begin{equation}
 V_Q = - \bl. \frac{1}{2} \bl( 1 - \frac{4}{\pi^2 Z^2 
               | H^{(2)}_0 (Z)|^4} \br),     
\label{QH} 
\ \ \ {\rm with}\ \ \  
  Z = \sqrt{-Z_+ \widetilde{Z}_-} = \biggl|\sqrt{V} 
    -2m+\frac{Q^2}{\sqrt{V}}-\frac{\lambda}{3}V^{3/2}\biggr|.  
\label{ZV}
\end{equation}
The behavior in the flat region $(Z \gg 1)$ and near 
the horizons$(Z \approx 0)$ is approximately 
\bea
V_Q \approx 
\left\{
\begin{array}{lr}
\displaystyle   0
& \ \ \ {\rm for}\ \ \ \ Z \gg 1 \\
\\
\displaystyle   \frac{1}{8 Z^2 (ln Z)^4}      
& \ \ \ {\rm for}\ \ \ \ Z \approx 0.
\end{array}   
\right.
\eea   
$V_Q$ approaches to zero at infinity, 
where the amplitude ${\sf R}$ becomes a constant. 
Near the horizon, on the other hand, $V_Q$ takes a infinite value.

Using the $U-\sqrt{V}$ relation (\ref{UVrel}), the remaining 
relation $T-\sqrt{V}$ is obtained in the integral form :
\begin{equation}
T=\frac{\pi}{2}\int Z | H^{(2)}_0 (Z)|^2 d\sqrt{V}.
\label{TV}
\end{equation}
We can show that the $T-\sqrt{V}$ relation approaches to the classical 
relation $T=\sqrt{V}$ in the semiclassical region.
We estimate the $T-\sqrt{V}$ relation (\ref{TV}) near the horizons as
\bea
T-T_0 \approx 
\left\{
\begin{array}{lr}
\displaystyle   \sqrt{V}
& \ \ \ \ {\rm for}\ \ \ \ Z \gg 1 \\
\\
\displaystyle   \frac12 (Z lnZ)^2 \sqrt{V}     
& \ \ \ {\rm for}\ \ \ \ Z \approx 0,
\end{array}   
\right.
\eea   
where $T_0$ is an integration constant.
The numerical estimation of Eq.(\ref{TV}) is shown in Fig.2.
We fixed $T=0$ at $\sqrt{V}=0$ and carried out 
the integral in Eq.(\ref{TV}). 
The rate of change of $dT/d\sqrt{V}$ of the dBB trajectory 
is diminishing as approaching 
to the horizons $\sqrt{V}=\sqrt{V_1},~\sqrt{V_2}$ and $\sqrt{V_3}~(Z=0)$. 
Near the horizons, $T$ of the quantum trajectory shows 
flat behavior and takes a finite value.
We connect the oppssite sides of the horizon smoothly 
on the dBB trajectory.
Fig.2 shows that the radial coordinate $T$ of(at) the horizons 
of the quantum black hole is smaller than that of the classical 
one, while the quantum surface area is identical with
the classical one ($4\pi V_1, 4\pi V_2$ and $4\pi V_3$).

Now we consider apparent horizons.  
The horizon is characterized by the expansions for the ingoing and 
outgoing null rays $\theta_-$ and $\theta_+$. 
Classically the equation $\theta_+=0$ expresses 
the apparent horizon. Therefore it has a local meaning, 
while the event horizons which exist at the null surface $U=0$ 
has a global meaning. 
In our black hole model based on the canonical approach, 
we have introduced the black hole mass as a dynamical variable on 
canonical data. 
The black hole mass have a classical relation to
the expansion $\theta_-\theta_+$:
\begin{equation}
 \theta_-\theta_+ = UV^{-1}(\dot{\sqrt{V}})^2 
   = -\frac{1}{4V}\biggl(1+\frac{{\it \Pi}_A^2}{V}-\frac{\lambda}{3}V
      -\frac{2M}{\sqrt{V}} \biggr).    
\end{equation}
In classical level the apparent horizon $\theta_-\theta_+=0$ 
agrees with the event horizon 
through the relation Eq.(\ref{u}).
In quantum level, the apparent and the event horizon are may be allowed 
not to coincide by quantum fluctuations.
Following ref. \cite{Nakamura93} 
we define the apparent horizons in quantum level by
$\Psi^{\ast}~ \hat{\theta}_+ \hat{\theta}_- \Psi = 0$.    
The mass operator $\widehat{M}$ and 
the electromagnetic momenta $\widehat{{\it \Pi}_A}$ 
in this definition are reduced to the eigen values $m$ and $Q$ 
respectively, and therefore $U=0$ must be satisfied on the 
quantum apparent horizons. 
The quantum apparent horizon agrees with the quantum event horizon
obtained by the dBB calculus. 
Thereby the separation of the event and the apparent horizon 
is not found from the dBB point of view.

In order to investigate the property of the horizon, 
we also calculate the radial motion of the light ray.
Here we treat the light ray as a classical object 
in the quantum background geometry.     
The coordinate of the light ray on the dBB trajectory is  
\begin{equation}
R =  \pm \frac{\pi}{2} ~\int \sqrt{V}~|H^{(2)}_0 (Z)|^2 d\sqrt{V} \\ 
\ \ \ {\rm with}\ \ \ \\
   Z=\biggl|\sqrt{V}-2m+\frac{Q^2}{\sqrt{V}}
           -\frac{\lambda}{3}V^{3/2}\biggr|.
\label{RV}
\end{equation}
For comparison, the light ray in the classical 
Reissner-Nordstr\"{o}m-de Sitter black hole is
\begin{equation}
 R_{\rm cl} = \pm \int \biggl(1-\frac{2m}{\sqrt{V}}+\frac{Q^2}{V}
         -\frac{\lambda}{3}V \biggr)^{-1} d \sqrt{V} 
\hspace{0.3cm}{\rm and}\hspace{0.3cm}\sqrt{V}=T.   
\end{equation}
Using Eq.(\ref{TV}), the approximate behavior near the horizon 
$Z \approx 0$ of Eq.(\ref{RV}) is estimated as 
\bea
R -R_0 \simeq
\left\{
\begin{array}{lr}
\displaystyle R_{\rm cl}\ 
& \ \ \ \ {\rm for} \ \ \ \ Z \gg 1 \\ 
\\
\displaystyle \mp \frac{\pi}{2}\sqrt{V} Z(ln Z)^2 
     & \ \ \ {\rm for} \ \ \ \ Z \approx 0 ,
\label{appRV}     
\end{array}   
\right.
\eea
where $R_0$ denotes an integration constant.
In Fig.3, the numerical estimation of Eq.(\ref{RV}) is shown. 
The integration constant $R_0$ is fixed at the origin $(T=0)$ and 
the infinity $(T=\infty)$ in order that the light ray 
on the quantum trajectory coincides with that on the classical 
trajectory.
The light ray forms a cups and reaches the horizons at finite $R$.        
%
\section{Summary and Discussion}
We studied the quantum geometrodynamics of the vacuum spacetime with 
spherical symmetry through a device of solving 
the Wheeler-DeWitt(WD) equation in the mini-superspace. 
We estimate the wave function from a viewpoint of the 
de Broglie-Bohm(dBB) interpretation. 
The distinctive feature of the dBB approach based 
on well defined quantum trajectories is that we can quantitatively 
estimate all quantum effects including the gravitation, 
even if the norm of the wave function cannot be defined. 
In a region where all quantum effects vanish, these trajectories are 
approach to classical ones. In this manner, we can always resolve  
the quantum theory into the (semi)classical one without having to 
assume an external observer and having to invoke the collapse 
of the wave function. In particular these properties are agreeable 
to quantum cosmology. 

In our calculation, we do not address a boundary contribution for the 
Hamiltonian which gives the ADM energy. 
For the system of coordinates on the 
hypersurface $\Sigma$ which is asymptotically Cartesian, 
the ADM energy of a black hole observed at the right and 
the left infinity is its black hole mass \cite{Kuchar94}.   
J. M\"{a}kel\"{a} and P. Repo studied the dynamical aspects of 
the quantum Reissner-Nordstr\"{o}m black hole \cite{JMPR}.    
In their model, the ADM mass and the electric 
charge spectra of the black hole are discrete. 
Moreover their semiclassical analysis showed 
that the horizon area spectrum of black holes 
is closely related to the Bekensteins's proposal. 
Our calculation shows that the quantum potential becomes zero at infinity. 
If there are the boundary terms, the quantum potential 
at the boundary will be significant\cite{Alwis94}. 

York introduced the idea of the quantum ergosphere that 
the apparent and the event horizon are separated quantum mechanically 
\cite{York83} and then 
Nakamura et al proposed that the gravitational fluctuation 
spontaneously induces the phenomena.
Our analysis shows that the mass eigen state does not distinguish 
the two horizons.
Rather, the quantum effect of the gravitational field makes 
the photon propagating along the radial coordinate reach 
the horizon within a finite time $R$. This may indicate that 
the causal connection between the opposite sides of 
the horizon is tied more strongly than in the case of the 
classical black hole. 
We can also propose that the quantum effect enhances 
the Hawking radiation. 
Further analysis is needed  to judge these suggestions.



\begin{figure}
\epsfxsize=7cm \epsfysize=5cm
\centerline{\epsfbox{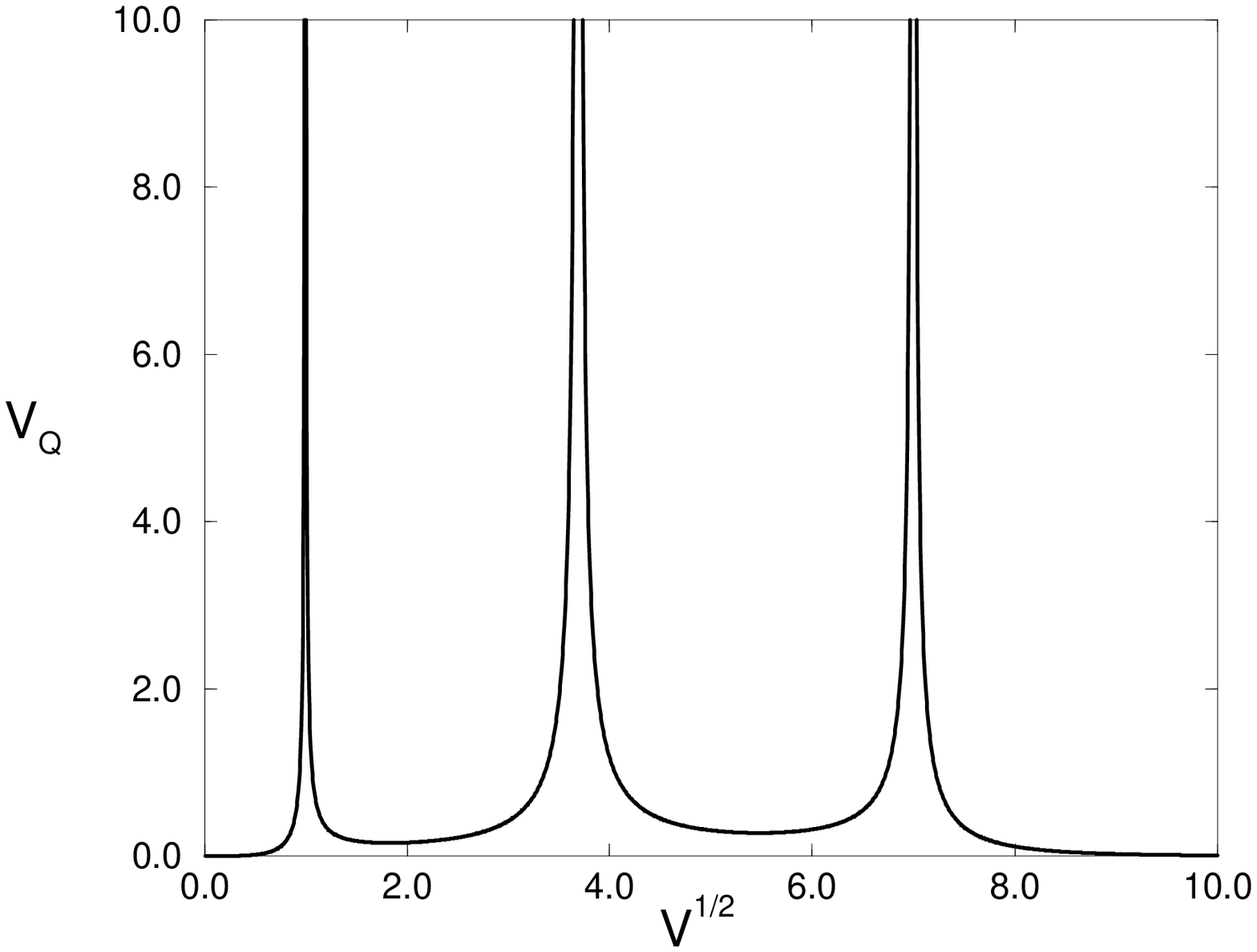}}
\caption
{The quantum potential for Reissner-Nordstr{\"o}m-de Sitter space. 
There are three singular points $\sqrt{V}=\sqrt{V_1},~\sqrt{V_2}$ 
and $\sqrt{V_3}$ for which $Z=0$. 
On each horizon, the quantum potential diverges 
and takes a positive infinite value.}
\label{Fig:QuantumPotential} 
\end{figure}

\begin{figure}
\epsfxsize=7cm \epsfysize=5cm
\centerline{\epsfbox{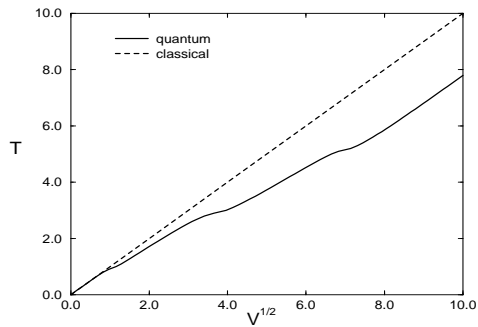}}
\caption
{The $T-\sqrt{V}$ relation is shown. The classical relation is denoted 
by the dashed line and the quantum one by the solid line. 
The rate of change $dT/d\sqrt{V}$ of the dBB trajectory 
is diminishing as approaching to the horizons 
$\sqrt{V}=\sqrt{V_1},~\sqrt{V_2}$ and $\sqrt{V_3}~(Z=0)$. 
$T$ shows flat behavior and takes a finite value on the horizons. 
}
\label{Fig:vtRelation} 
\end{figure}

\begin{figure}
\epsfxsize=7cm \epsfysize=5cm
\centerline{\epsfbox{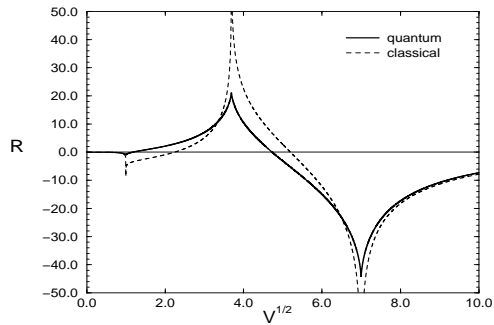}}
\caption
{The light ray on the dBB trajectory is shown by the solid line. 
The dashed line is the light ray in the classical geometry. 
The light ray on the dBB trajectory forms a cups and reaches 
the horizons at finite $R$.       
}
\label{Fig:LightRay} 
\end{figure}



\begin{references}

\bibitem{Dirac64}
P. A. M. Dirac, {\it Lectures on Quantum Mechanics} 
(Yeshiva University, New York, 1964).

\bibitem{ADM62}
R. Arnowitt, S. Deser, and C. W. Misner, 
in {\it Gravitation: An Introduction to Current Research}, 
edited by L. Witten (Wiley, New York, 1962).

\bibitem{Wheeler}
J. A. Wheeler, {\it in Batelle Rencontres: 1967 Lectures in 
Mathematics and Physics}, edited by C. DeWitt and J. A. Wheeler 
(Benjamin, New York, 1968) pp .   

\bibitem{DeWitt67}
B. S. Dewitt, Phys. Rev. {\bf 160}, 1113 (1967). 

\bibitem{Kuchar94}
K. V. Kucha$\check{\rm r}$, Phys. Rev. {\bf D50}, 3961 (1994).

\bibitem{Nakamura93}
 K. Nakamura, S. Konno, Y. Oshiro, A. Tomimatsu,  Prog. Theor. Phys.  
 {\bf{90}}, 861 (1993).

\bibitem{Hawking75}
S. W. Hawking, Comm. Math. Phys.{\bf 43}, 199 (1975).

\bibitem{Gibbons-Hawking}
G. W. Gibbons and S. W. Hawking, Phys. Rev. {\bf D15}, 2738 (1976).

\bibitem{KKTY}
M. Kenmoku, H. Kubotani, E. Takasugi, Y. Yamazaki, Phys Rev {\bf D57}, (1997).

\bibitem{deBroglie}
L. de Broglie, Tentative d'interpretation causale et non-lineaire 
de la mechanique ondulaire (Gauthier-Villars, Paris, 1956).

\bibitem{Bohm52}
D. Bohm, Phys. Rev. {\bf 85}, 166, 180 (1952).  

\bibitem{Bell87}
J. S. Bell, {\it Speakable and Unspeakable in Quantum Mechanics} 
(Cambridge University Press, Cambridge, 1987).

\bibitem{Halliwell87}
J. J. Halliwell, Phys. Rev. {\bf D36}, 3626 (1987).


\bibitem{JMPR}
J. M\"{a}kel\"{a}, and P. Repo, Phys. Rev. {\bf D57}, 4899 (1998). 

\bibitem{Alwis94}
S. P. Alwis, D. A. MacIntire, Phys. Rev. {\bf D50}, 5164 (1994).

\bibitem{York83}
James W. York, Jr., Phys. Rev. {\bf D28}, 2929 (1983).

\end{references}
\end{document}